\documentclass[prb,twocolumn,amsmath,amssymb,floatfix]{revtex4}
\usepackage[dvips]{graphicx}
\usepackage{color,array,dcolumn}
\usepackage{amsmath}
\usepackage{amssymb}

\begin{document}

\title{Ballistic atomic transport in narrow carbon nanotubes}

\author{Alberto Ambrosetti$^1$\footnote[1]{Corresponding author. Email:alberto.ambrosetti@unipd.it}, Pier Luigi Silvestrelli$^1$, John F. Dobson$^2$, Luca Salasnich$^{1,3,4}$}

\affiliation{
$^1$Dipartimento di Fisica e Astronomia, Universit\`{a} degli Studi di Padova, via Marzolo 8, \textsl{35131}, Padova, Italy\\
$^2$School of Environment and Science, Griffith University , \textsl{4111}Nathan Queensland, Australia\\
$^3$Istituto Nazionale di Ottica del Consiglio Nazionale delle Ricerche, Unit\`a di Sesto Fiorentino,
via Carrara 1, 50019 Sesto Fiorentino (Firenze), Italy \\
$^4$Istituto Nazionale di Fisica Nucleare, Sezione di Padova, via Marzolo 8, \textsl{35131}, Padova, Italy
}


\begin{abstract}
\date{\today}
Friction forces are conventionally modeled via semi-classical theories, that associate energy dissipation with newtonian motion on corrugated interface potentials.
This consolidated approach is challenged at the nanoscale by observation of nearly-unimpeded water flow in narrow carbon nanotubes (CNTs), in spite of non-vanishing energy corrugations.
Here we go beyond the standard newtonian perspective, adopting a quantum mechanical description of $^4$He flow through narrow CNTs. Building upon our Bloch-wave dynamics
[PRL {\bf 131}, 206301 (2023)] we explore realistic flow conditions, including non-negligible interface interactions, finite temperatures, and imperfect CNTs.
At $T$=0 K we found that $^4$He waves can propagate through ideally periodic, corrugated interface potentials with no friction:
below a critical velocity regulated by interface corrugations, energy loss by emission of plasmon and phonon quanta is forbidden.
Introducing realistic impurities/defects one still finds very large mean free paths that can exceed the $\mu$m scale,
while thermal phonons and plasmons yield even lower scattering rates.
This establishes the unexpected emergence of ballistic wave-like transport in narrow CNTs within realistic nanoscale devices, and demonstrates the intrinsic quantumness of nanoscale interfaces.
\end{abstract}

\maketitle

The concept of ballistic transport (BT) is conventionally associated with the near-unimpeded flow of charge carriers~\cite{todorov} through quasi one-dimensional (1D) crystalline systems, such as carbon nanotubes~\cite{ando2,poncharal} (CNT) or nanowires~\cite{chan,sakr}.
In normal conductors, the collision of charge carriers on impurities/defects or thermal ionic displacements
implies finite resistivity, and prevents unimpeded flow.
However, when the mean free path (i.e. the average length a particle travels between subsequent collisions) largely exceeds
the longest dimension of the medium, BT becomes possible and the measured resistivity can closely approach zero. Two main consequences emerge: {\it i)}  the rise of virtually non-dissipative electric
currents, and {\it ii)} the appearance of {\it coherence} phenomena~\cite{kong,bohm} related to the non-local quantum nature of the
nearly-unperturbed flowing particles.

As recently demonstrated by unimpeded phonon dynamics~\cite{maire,nomura}, BT mechanisms are not necessarily restricted to charge carriers.
In principle, generalization of BT to alternative quasi-particles under suitable conditions could be equally viable -- although such phenomena remain so far largely unexplored.

From the experimental side, nanofluidics measurements conducted at room temperature have recently reported exceptionally-high water flow rates~\cite{majumder,holt,whitby,secchi} through
narrow carbon nanotubes. Measured permeabilities exceed classical fluid-mechanics/molecular dynamics predictions~\cite{kannam} by up to 4 orders of magnitude,
implying a dramatic reduction of friction
forces. Very steep permeability growth~\cite{secchi,michaelides} is also observed when the CNT radius is reduced towards the nm scale, in spite of the higher surface-to-volume ratio.
The failure of (semi-)classical approaches suggests~\cite{bocquet,michaelides2,sokoloff1,sokoloff2,coquinot} deep quantum mechanical transport. For instance,
momentum-flow tunnelling~\cite{coquinot2} and quantum fluid friction~\cite{bocquet}, were recently proposed in analogy
to electronic tunnelling, and to electronic friction.
On the other hand, the 1D character of the effect also introduces a clear parallelism with electronic BT.
The question, then, is whether BT can actually extend to atomic/molecular flow or not.

Starting from a quantum description of confined atomic flow built upon our Bloch-wave dynamics
model -a model which we recently exploited~\cite{prl} to describe He flow at $T$=0 K in ideally periodic CNT's- we will capture realistic effects such as corrugated interface potentials, impurities and thermal
excitations. As a main result, we will confirm the viability of ballistic atomic transport (denoted hereafter as BAT).
So far, unimpeded mass flow was essentially associated with superfluidity and limited to ultra-low temperatures.
BAT could extend near-frictionless flow to room conditions, opening the way to innovative nanofluidics technologies
such as ultra-efficient gas filtration, self-sustained water purification~\cite{filter}  and non-destructive injection in cellular membranes.

\section{Quantum Bloch-wave model for confined flow}\label{sec2}

To explore the viability of BAT and to unambiguously identify the relevant physical ingredients, we will treat a simplified problem,
considering the flow of individual, weakly interacting $^4$He atoms through a narrow and ideally periodic (5,5) CNT (with diameter $d=6.82$ \AA).
Our model will be valid in the limit of low $^4$He density and will be eventually transferable to alternative chemical moieties.
$^4$He is considered here in place of water due to its isotropic character, that avoids  permanent dipole moments.
$^4$He also has a very small size, and it interacts weakly both with CNT and with other $^4$He atoms, justifying the
adoption of a single-atom approach in the low density limit.
Considering that the $^4$He-$^4$He interaction energy amounts~\cite{hodges} to $\sim$1 meV at a separation of $\sim$3 \AA, non-linear effects due
to multiple flowing atoms can be neglected when the $^4$He-$^4$He distance exceeds $\sim$6 \AA.
The key ingredients for the onset of BAT are all retained by our model. Laterally confined
$^4$He performs an effective 1D motion~\cite{prl,jcp} along the CNT axis $\hat{x}$. Due to the interaction with the CNT walls $^4$He is subject to an effective interface
potential $V_{\rm eff}(x)$, periodically repeated along $x$. The CNT unit cell has length $L=$2.46 \AA\,  and $m_{\rm He}$ indicates the $^4$He mass.

Tribology problems are conventionally addressed in literature~\cite{carkner,tosatti,righi,zhong,wgao,tocci,waterslide} from a semi-classical Newtonian perspective, associating friction forces
with the existence of corrugated interface
potentials that must be overcome upon relative motion. However, even when moving at a very fast speed of $\sim$200 m/s, a $^4$He atom is already associated to a De Broglie
wavelength that exceeds the CNT unit cell size. Hence, $^4$He cannot perceive $V_{\rm eff}(x)$ in a classical Newtonian sense.

To account for the wave-like nature of the flowing $^4$He,
we exploit an effective quantum Hamiltonian~\cite{prl} (atomic units will be exploited hereafter):
\begin{equation}
        H_{\rm eff} = -\frac{1}{2m_{\rm He}} \frac{d^2}{dx^2} + V_{\rm eff}(x, \mathbf{R})\,.
        \label{model1d}
\end{equation}
Here $V_{\rm eff}$ depends on the $^4$He coordinate $x$ and carries additional dependence on the coordinates $\mathbf{R}$ relative to the CNT C atoms.
A direct estimate of $V_{\rm eff}$ for the ideally periodic CNT is obtained by first-principle density functional
theory (DFT -- see details below). All C atoms were held in their equilibrium configuration ($\mathbf{R}_{\rm eq}$). In general, one has
$V_{\rm eff}(x,\mathbf{R}_{\rm eq})=\sum_{J=1}^{\infty} A_{J} \cos {(K_{J} x)} + B_{J} \sin {(K_{J} x)}$ where $K_{J}=2J\pi/L$ (phonons were later introduced perturbatively to capture CNT
atomic motions).
Laterally, the $^4$He atom is tightly confined within the CNT.
Hence, transversal motion is effectively factorized in the wavefunction (more details will be provided below).
The summation in $V_{\rm eff}$ is dominated~\cite{jpcc,jcp} by the cosinusoidal term $J=2$. In fact $V_{\rm eff}$ is nearly cosinusoidal, with $A_2\sim 0.03$ meV.
Even smaller corrugation amplitude ($A_2\lesssim0.01$ meV) is
found in the slightly larger (7,7) CNT (both armchair CNTs have metallic nature), whereas larger corrugation ($A=$0.76 meV) is associated with the finite-gap (8,0) CNT, in spite of the
slightly larger diameter ($d$=0.76 \AA).

Diagonalization of $H_{\rm eff}$ yields a set of single-particle Bloch wavefunctions $\psi_{q,n}(x)$ with corresponding
energy levels $E_{q,n}$ (having band index $n$ and wave vector $q$ within the first Brillouin Zone --1BZ--, i.e. $-\pi/L<q<\pi/L$).
These build up the band structure of flowing $^4$He (see Fig.~\ref{bands}). If $V_{\rm eff}$ only contains the term $A_{2} \cos{(K_{2} x)}$,
the periodicity of the system becomes compatible with $L/2$. This formally enlarges the BZ to twice the standard CNT 1BZ size.
After folding $^4$He bands into the standard 1BZ, band gaps appear
at $q=0$ (see Fig.~\ref{bands}). When additional terms are introduced in $V_{\rm eff}$, small gaps emerge also at the
1BZ edges. Transversal motion is initially neglected, being associated with larger energy scales with respect to low-lying bands (see below).

To describe friction one must now explicitly account for the quantum mechanical excitations in the CNT.
In fact, $^4$He could perform quantum mechanical scattering, and transfer part of its energy~\cite{prl,jcp} to the low-lying CNT quasi-particle modes,
undergoing effective dissipation.
The relevant low-lying quantum modes here are the acoustic CNT phonons (quantum vibrations of the atomic lattice), and plasmons (collective electron excitations).
At low momenta $p$, these modes exhibit quasi-linear energy dispersion. For instance, the {\it softest} transverse acoustic (TA) CNT phonon has~\cite{dresselhaus,zhang}
a dispersion relation $\omega_{\rm TA}(p) \sim v_{\rm TA} |p|$ (with $v_{\rm TA}\sim 4.5\times10^{-3} a.u.$ in the (5,5) CNT).
Other phonon modes and plasmons~\cite{mbdc,jcp} are associated with higher velocities.

\section{Role of periodic interface potentials at $T$=0 K}

\begin{figure}[bt!]
\includegraphics[width=8.5cm]{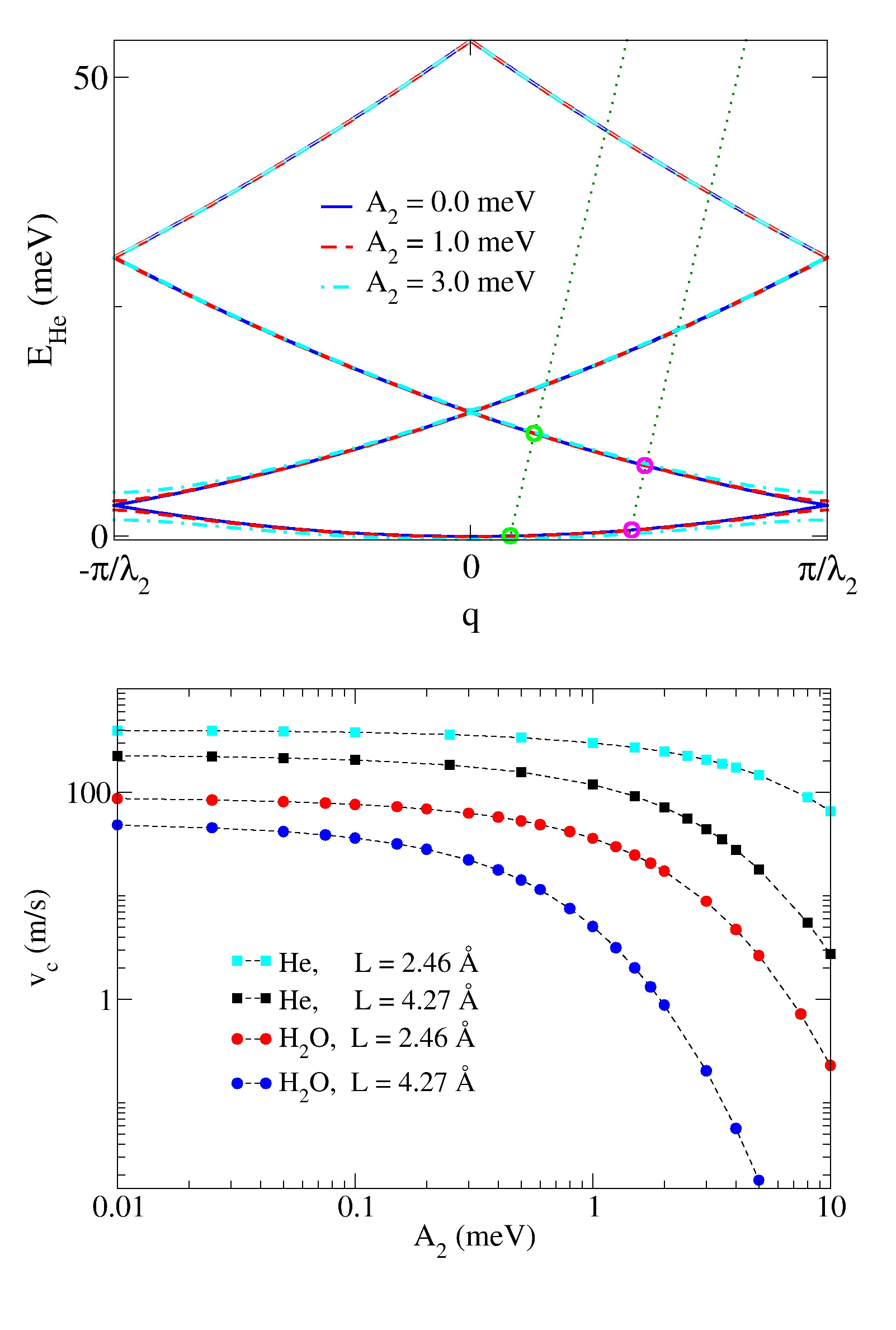}
        \caption{{\bf a)} Band structure (longitudinal, i.e. related to motion along $x$) for a single $^4$He atom in periodic
        potentials due to the
        interaction with CNT walls. At $A_2=0$ $^4$He behaves as a free particle, with parabolic bands.
        Extremely small energy gaps emerge at $A_2=0.03$ meV, and the deviation from parabolic bands
        is hardly detectable. When $A_2$ becomes larger, gaps become more visible and bands gradually tend to flatten.
        The unit cell length is fixed to $L=2.46$ \AA, compatibly with the (5,5) CNTs.
        {\bf b)} Critical velocity $v_c$ --i.e.  maximum slope of the ground state band--for
        $^4$He immersed in $V_{\rm eff}$ with variable amplitude $A_2$. Two different periodicities $L$
        (compatible with zigzag and armchair CNTs) are taken into consideration.
        }
\label{bands}
\end{figure}

We begin considering an ideally periodic CNT at $T$=0 K (our Bloch-wave model~\cite{prl}, will be improved including non-negligible $V_{\rm eff}$).
Under these conditions, a $^4$He atom with initial momentum $q$ and energy $E_{q,n}$ could only
scatter against the CNT exciting phonons or plasmons from the vacuum (both modes will have momentum $p$).
After scattering, $^4$He acquires momentum $q'$ and
energy $E_{q',n'}$. The energy change in the presence of CNT atomic displacements $\delta R_{i,l_c}$
($i$ indicates an atomic Cartesian coordinate within the unit cell, while $l_c$ labels the unit cell replica) when $^4$He occupies the coordinate $x$
is
$\Delta U=-\sum_{i}\sum_{l_c=1}^N F_{i,l_c}(x) \delta R_{i,l_c}$. We will let $N$  finally tend to infinity. Here $F_{i,l_c}$ is the force acting on C atoms due to
the presence of $^4$He, and it is obtained from the interface potential as $F_{i,l_c}=-\partial_{R_{i,l_c}} V_{\rm eff}$.
This expansion introduces a perturbative quantum mechanical treatment of the  nuclear motion.
The scattering rate for the leading one-phonon (of type $j$) process is described by Fermi's golden rule:
\begin{eqnarray}
        \Gamma^{\rm ideal} =2\pi \vert  \langle 1^j_{p} | \langle \psi_{q',n'} | \Delta U  | \psi_{q,n} \rangle | 0 \rangle \vert^2 \nonumber \\
        \times \delta(E_{q,n}-E_{q',n'}-\omega_j(p))= \nonumber \\
        =2\pi \vert \sum_{i,l_c} \langle \psi_{q',n'} | -F_{i,l_c}(x) | \psi_{q,n} \rangle \langle 1^j_{p} | \delta R_{i,l_c}  | 0 \rangle \vert^2 \nonumber \\
        \times \delta(E_{q,n}-E_{q',n'}-\omega_j(p))\,.
        \label{conserv}
\end{eqnarray}
Calculations are reported in full detail in the Supplemental Material~\cite{SM}.
After suitable rotation to collective vibrational coordinates, $\delta R_{i,l_c}$ can be recast as a linear combination of phonon creation/annihilation operators, and thus
accounts for single phonon absorption/emission.
As previously noted, at $T$=0 K the phonon modes are in their ground state, so absorption of a phonon quantum is impossible: only phonon emission
(i.e. excitation of a vibrational quantum from the vacuum) can occur, and we will now show that this process also has limitations.
Considering for instance the excitation of TA phonons (extension to other phonons or plasmons is straightforward), integration of Eq.~\eqref{conserv}
implies the following two-fold conservation laws:
\begin{eqnarray}
        \left\{
                \begin{array}{l}
E_{q,n}=E_{q',n'}+\omega_{\rm TA}(p)\,,  \\
q=q'+p+lK_1\,,
                \end{array}
                \right.
\label{system}
\end{eqnarray}
The conservation of crystal momentum (i.e. momentum conservation up to integers multiples $l$ of $K_1$) is a consequence of the discrete translational
invariance of the CNT. Considering the near-linear dispersion of TA phonons, Eq.~\eqref{system} can be satisfied only if there exists a unique phonon
dispersion line that intersects $^4$He bands at $\left(q,E_{q,n}\right)$ and $\left(q',E_{q',n'}\right)$ at the same time (see Fig.~\ref{bands}).
The double intersection automatically ensures conservation of energy and crystal momentum. Observing the band structure, it is obvious that the high slope
of the phonon dispersion (determined by $v_{\rm TA}$) is incompatible with a double intersection within the ground-state band ($n=0$).
As a consequence, at $T$=0 K
scattering is possible only when $^4$He initially occupies an upper band ($n>0$) and decays onto a lower band. However, no scattering can occur if $^4$He
initially occupies the ground-state band ($n=0$): in this case no phonon can be excited conserving energy and crystal momentum.

This kinematic constraint, -- formally analogous to Landau's superfluidity~\cite{prl,landau,superf2} criterion -- quantum-mechanically protects the particle from scattering.
Virtually frictionless particle flow is thus possible at $T$=0 K in perfect CNTs, as long as the particle sits in the ground state band.

We now explicitly consider the role of non-vanishing energy corrugations.
Since the flowing particle has wave-like character, its speed can be identified with the group velocity  $v_n(q)=\partial E_{q,n}/\partial q$.
The restriction to the ground state band can thus be reformulated in terms of a velocity constraint. In fact, as long as the particle speed $v$ does not
exceed a critical velocity $v_{\rm c}$ such that
\begin{equation}
v_{\rm c}={\rm Max}_{\{q\in BZ\}} \left(\frac{\partial E_{q,0}}{\partial q}\right)\,,
\end{equation}
unimpeded flow is possible.

In the limit of vanishing $V_{\rm eff}$, $^4$He bands become parabolic, and $v_{\rm c}=K_1/2m_{\rm He}$. Instead, when the interface potential has non-negligible
corrugation, $^4$He bands tend to flatten, opening larger gaps. The consequence is a reduction of the critical velocity $v_{\rm c}$, which limits the
velocity range of the unimpeded flow. In the limit
of very large corrugation $v_{\rm c}$ finally tends to zero, due to $^4$He confinement. This provides a clear connection between quantum
mechanics and semi-classical tribology approaches,
where  higher interface energy corrugations are associated with larger friction forces.

Large $v_{\rm c}$ (exceeding 10 m/s) can persist when the potential corrugation reaches a few meV.
Frictionless particle flow at $T$=0 K is thus expected to be possible also on alternative weakly-coupled interfaces.
However, when  $v > v_{\rm c}$ occupation of higher-energy bands becomes unavoidable and scattering is permitted. Finite friction forces
can potentially arise.

In spite of the evident analogy with Landau's superfluidity criterion
and the rise of frictionless flow, no superfluidity is present here.
In fact, superfluidity emerges as a many-body effect, while only a
single He atom is considered here. In a conventional superfluid
flowing particles and collective quasi-particle modes are both part of
the superfluid itself. Instead, here one has a net distinction between
the flowing particle (He), and the collective excitations (living in the CNT).

We observe for completeness that even larger velocities (reaching 82 m/s)
were found~\cite{Codiff} for Co diffusion on the external surface of CNTs.

\section{Scattering by impurities}
A key assumption introduced so far was the ideal periodicity of the CNT lattice, which ensures conservation of crystal momentum.
However, a few impurities may be present in CNTs, and could in principle introduce finite friction forces.
Among these, the most frequent are Stone-Wales~\cite{SW} lattice defects - which still exhibit extremely low concentrations, i.e. up to
$10^{-15}$ in chemical vapor deposition (CVD)-grown CNTs. The attractive vdW energy of $^4$He at closest approach on Stone-Wales defects is of the order~\cite{SW,SW2,jpcc11}
of $\sim10^{-2}$ eV. Much more abundant disturbances can be determined by absorbates on the outer CNT wall.
Weak physical absorption of environmental gaseous moieties commonly takes place on CNTs, governed by weak non-covalent interactions.
The effect of outer absorbates on the inner $^4$He is due to weak vdW interactions, further screened by the CNT walls.
Even neglecting screening, the binding energy of $^4$He due to an outer NH$_3$ molecule (one of the most stable~\cite{leenaerts} physisorbate moieties)
amounts to only $\sim 0.05$ meV, as estimated through the Tkatchenko-Scheffler~\cite{TS} approach.

We will determine hereafter the scattering rate for a $^4$He atom flowing through a (5,5) CNT on the aforementioned impurities.
Given the very small $V_{\rm eff}$ corrugation we approximate longitudinal $^4$He wavefunctions as single plane waves associated with
parabolic energy bands (the corresponding band index $n$ could thus be dropped by letting momentum exceed the BZ).
As previously mentioned, the lateral $^4$He confinement in a (5,5) CNT within the relevant energy scales is well approximated by a harmonic potential,
with frequency $\omega_{\rm c}\sim 23$ meV. Hence,  one can describe the single unperturbed $^4$He through the following states:
\begin{equation}
        \psi_{q,n_y,n_z}(x,y,z) = \frac{e^{iqx}}{\sqrt{L_{\infty}}} \phi_{n_y}(y) \phi_{n_z}(z)\,,
\end{equation}
where $\phi_{n}$ is the $n$-th 1D oscillator eigenstate relative to frequency $\omega_{\rm c}$. The normalization of the longitudinal plane wave is
imposed over a large simulation  box of length $L_{\infty}=NL$ (where presence of a single impurity is assumed).
We note that the finite lateral extension of the He wavefunction implies
that the interface potential should be averaged out along the transversal directions.
However, given the small width of the oscillator ground state, modest
renormalization (of the order of 0.5 meV) is expected.

Supposing $^4$He initially occupies the state $(q$,$n_{y}$,$n_{z})$, the scattering rate to a final state $(q'$,$n'_{y}$,$n'_{z})$, by an impurity potential $V_{\rm imp}$
is:
\begin{eqnarray}
        \Gamma^{\rm imp} = 
        2\pi \vert\langle \psi_{q',n'_y,n'_z} | V_{\rm imp} | \psi_{q,n_y,n_z}\rangle   \vert^2 \nonumber \\ 
        \times \delta(E_{(q,n_{y},n_{z})} - E_{(q,n'_{y},n'_{z})})  \,.  
\end{eqnarray}
We suppose the $^4$He atom initially sits in the lowest {\it longitudinal} band (i.e. $q\in$ 1BZ) and in the oscillator ground-state (at $T$=0 K).
Since the periodic interface potential is neglected (extension of the model is straightforward), the initial state will be labeled by ($q$,$n_x$=0,$n_y$=0).

Due to energy conservation,  $^4$He  can only backscatter, inverting its momentum from $q$ to $q'=-q$ and remaining in ($n_x$=0,$n_y$=0).
By definition,  impurities are not periodic. Hence, quasi-momentum is not conserved in the scattering.

Both SW impurities and adsorbates are modeled here through an effective 1D vdW potential (see Supplemental Material~\cite{SM}):
\begin{equation}
        V_{\rm imp,1D}(\mathbf{x})=-C_{6,{\rm eff}} \left(x^2+d^2\right)^{-3}\,.
\end{equation}
The parameter $V_0=-C_{6,{\rm eff}} d^{-6}$ is identified with the potential depth, while $d$ provides the effective radial distance of the source of the vdW potential.
The matrix element depends on the Fourier transform of $V_{\rm imp}$, which exhibits exponential decay with respect to $|q|$,
and implies suppression of $\Gamma^{\rm imp}$ at large momentum.

For nearly free particles, the mean free path can be finally estimated as
\begin{equation}
\lambda^{\rm imp}=(\Gamma^{\rm imp})^{-1}|q|/m_{\rm He},
\end{equation}
where $q$ is the initial momentum of the particle.

\begin{figure}
\includegraphics[width=8.5cm]{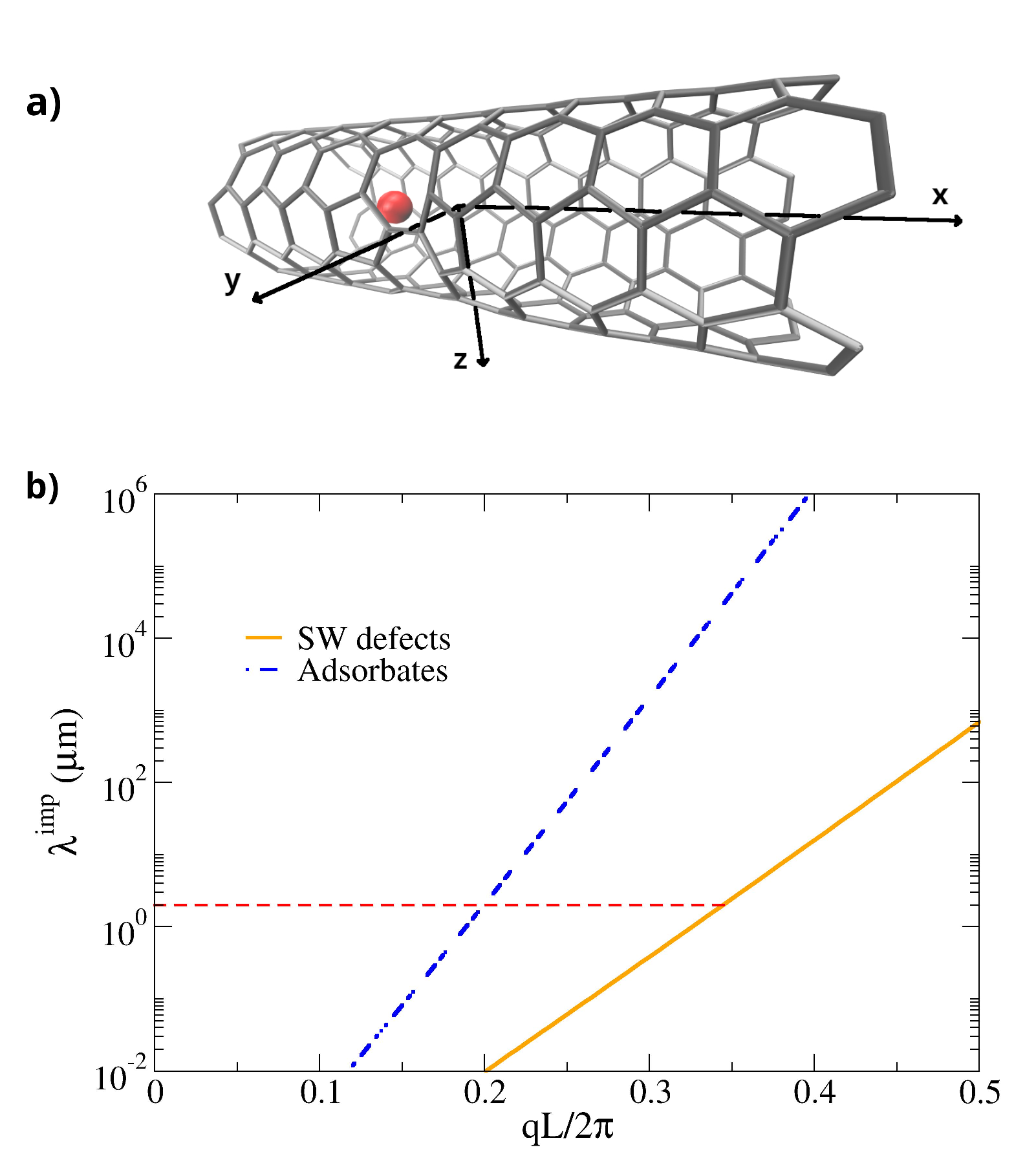}
        \caption{He scattering on CNT impurity. {\bf a)} Geometry of a Stone-Wales defect in the (5,5) CNT, as modeled in this work.
        The flowing He atom is depicted in red. Cartesian axes are reported for reference. The defect is located on the CNT wall at $(x=0,y=R,z=0)$.
        {\bf b)} Mean free path for impurity scattering, as a function of the initial momentum $q$.
        Initial oscillator states are set to the ground state for reference ($n_y=n_z=0$), and summation over possible final states is performed.
        Different potential depths ($V_0$=10 meV and 0.05 meV) and average distances between impurities
        ($\Delta$=2 $\mu$m and 5 \AA) are considered, depending on the defect type (SW or adsorbate, respectively).
        The present choices for $\Delta$ are conservative and can be significantly increased for both SW defects and adsorbates, with adequate fabrication
        techniques, controlling environmental properties or adopting concentrical CNTs with reduced spacing. Larger $\Delta$ values
        imply larger mean free paths.
        The dashed red line indicates the lower bound for $\lambda^{\rm imp}$ relative to SW defects. This corresponds to the
        average distance between defects adopted here (the value is prudential and can be significantly enhanced).
        }
\label{fig2}
\end{figure}

As shown in Fig.~\ref{fig2}, $\lambda^{\rm imp}$ becomes very large at high momentum (close to BZ edges), exceeding the $\mu$m scale,
and implying deep ballistic motion. In estimating the impact of SW defects we took an average separation $\Delta\sim2$ $\mu$m. This sets a lower bound to the mean free path at $\sim$2 $\mu$m. However, the extremely low defect concentrations that can be realized at the experimental level can significantly increase this length scale.
Concerning absorbates, we note that complete physisorption with a periodic pattern compatible with the CNT lattice
would simply renormalize $V_{\rm eff}$, without causing scattering.

Excitation of transversal oscillator modes could open additional scattering channels. However, these channels
are associated with large momentum variation, which implies strong suppression. 
The situation is expected to change in CNTs with large diameter, where transversal
oscillator modes are associated with lower energy scales and parabolic bands are
replicated at small energy intervals. In fact, the transversal component of the He
wavefunction can become ring-shaped in large CNTs, due to absorption on internal walls.
The weaker lateral confinement is then expected to reduce the corresponding excitation
energy scale.
In that case,  impurity scattering can become relevant due to low momentum transfer between neighboring band replicas.
The exponential dependence of the Fourier-transformed interaction on momentum transfer suggests fast variation of scattering rates
with respect to the CNT diameter. This implies a strong connection between BAT and near 1D confinement,
in analogy with electronic BT. We note in passing that water-flow experiments actually reveal extremely rapid permeability increase~\cite{bocquet}
when CNT diameters are decreased below the $\sim10$ nm scale.

\section{Scattering by thermal phonons}
So far, calculations were performed at zero temperature. However, thermal contributions could also potentially interfere with the particle flow
introducing effective dissipation. In fact,
the absorption of thermally excited phonons/plasmons could lead to inter-band transitions that can alter (or even invert) the velocity of the flowing particle.
Again, we will initially consider only phononic degrees of freedom, given the formal~\cite{jcp} analogy with plasmons (plasmons will be discussed later). At finite temperature, the average number of
excitations associated with the phononic frequency $\omega_p^j$ will be $n_p^j(T)=\frac{1}{e^{\omega_j(p)/K_B T}-1}$
where $K_B$ is Boltzmann's constant.
Extending Eq.~\eqref{conserv}, we will enable a $^4$He atom (with initial and final momenta $q$ and $q'$) to absorb energy from the CNT annihilating a
thermally excited phonon with momentum $p$. The corresponding matrix element, accounting for a change in the population of the $j$-th phonon
from  $n_p^j$ to $n_p^j-1$ is:
\begin{equation}
        -\sum_{i,l_c} \langle q' | F_{i,l_c} | q \rangle \langle n_{p}^j-1| \delta R_{i,l_c} | n_p^j \rangle \,. 
\end{equation}
Defining the Fourier transform $\tilde{f}_a(p)=\int_{-L_{\infty}/2}^{L_{\infty}/2}  F_{a,0}(x) e^{-ip x} dx$,
the transition rate $\Gamma^{\rm therm}$ can be computed from the above matrix elements in analogy with Eq.~\eqref{conserv}.
Conservation laws analogous to Eq.~\eqref{system} are enforced here, with the phonon being absorbed rather than emitted.

\begin{figure}
\includegraphics[width=8.5cm]{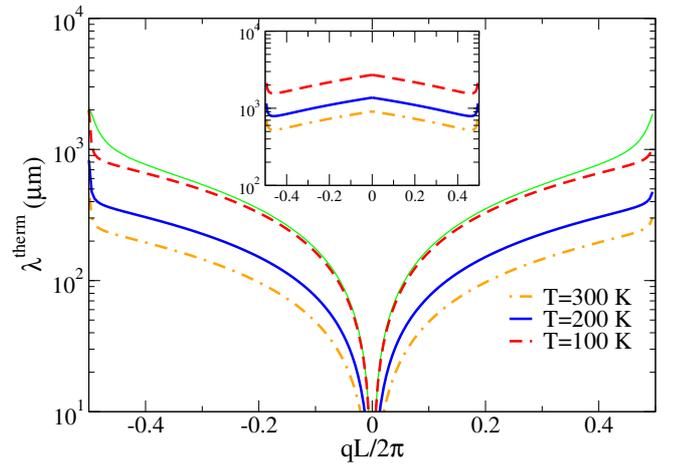}
        \caption{Mean free path for thermal scattering $\lambda^{\rm therm}$, involving single-phonon absorption as a function of
        the initial momentum $q$, at different temperatures $T$. Summation over available final states is performed.
        A very small band gap of 0.001 meV is assumed at BZ edges, compatibly with the CNT periodicity. Results show minor
        dependence on this band gap: the green line shows data for $T$=100 K, with a larger band gap of 0.1 meV (larger band gaps would further increase the mean free path).
        Inset: scattering rates relative to the transition between first-excited longitudinal band and ground state band, via phonon emission.
        }
\label{fig3}
\end{figure}

To numerically estimate $\Gamma^{\rm therm}$ for the case of He in a (5,5) CNT we observe that particle-CNT forces are dominated by non-covalent
vdW interactions.
This is clearly confirmed by first-principle calculations: long-range vdW corrections~\cite{d2} enhance He-(5,5)CNT interface corrugations by a factor of $\sim3$ with
respect to pure PBE (which only accounts for the shorter-range component of vdW forces). A standard pairwise vdW estimate based on the TS~\cite{TS} procedure
is adopted here (see Supplemental Material~\cite{SM}).
We also note that the presence of band gaps induced by the corrugated potential avoids divergence at the BZ edges ($\omega^j(p)$ will not vanish there).
As from Fig.~\ref{fig3} the mean free paths (estimated in analogy with $\lambda^{\rm imp}$) can exceed the $\mu$m scale even at $T$=300 K,
when $q$ is sufficiently large. As a consequence, BAT remains viable even at room temperature.

In addition, even the occurrence of rare one-phonon scattering events does not necessarily imply major flow
reduction. In fact,
phonon absorption can both decrease and increase the particle speed. Such process could substantially mitigate the net flow
reduction.

The flowing $^4$He atom is not treated here as a thermal particle. In fact, we assumed its initial
speed can be adequately controlled, and $^4$He can be easily cooled down before injection.
Considering the extremely large mean free paths, thermalization by contact with the CNT during
the flow can be excluded. In the absence of cooling, fractional occupation of excited modes could enable
scattering by inter-band transition and phonon emission. However, after decay to the ground state band, $^4$He will remain there, preserving
BAT on long time-scales.

Rates of transition between the first excited and the ground state longitudinal band by phonon emission are also easily
computed changing initial and final states in Eq.~\eqref{conserv}. Such process describes the scattering of $^4$He when its speed exceeds $v_c$.
As from Fig.~\ref{fig3} (inset), mean free paths remain very
large even in the supercritical regime, suggesting moderate increase of friction forces with respect to the initial velocity.

A slightly more complex description is required for $^4$He-plasmon~\cite{dobson,perez,plasm} scattering (details are reported in the Supplemental Material~\cite{SM}). Due
to the absence  of permanent multipoles in $^4$He, single-plasmon emission/absorption is not permitted, and scattering
processes involve plasmon pairs and second-order
perturbative contributions. At finite $T$ this potentially implies possible intraband transitions even below $v_{\rm c}$. In practice,
the mean free paths associated with such (leading) process largely exceed those from phonon scattering, and can be neglected (see Supplemental Material~\cite{SM}).

We finally observe that under 1D confinement the attractive potentials generated by impurities are associated to bound states.
In principle, unbound $^4$He could decay onto a bound state by phonon emission. Explicit calculations (see Supplemental Material~\cite{SM})
show that at sufficiently high speed even this type of decay becomes extremely rare, due to the tendency of the system to conserve linear momentum.

\section{Discussion}

In the limit of $T$=0 K
and ideal CNT lattice, individual $^4$He atoms can virtually flow through the CNT without friction, at variance with semi-classical models.
A quantum mechanical kinematic constraint, formally analogous to Landau's superfluidity~\cite{landau} criterion protects $^4$He from scattering
as long as a critical velocity $v_{\rm c}$ is not exceeded.
Similar constraints also apply to neutron scattering experiments~\cite{ashcroft}, and their well-established experimental observation confirms the robustness of
the mechanism.
Ideally periodic CNT-$^4$He interface interactions do not necessarily introduce finite friction forces: they rather cause renormalization of the critical velocity $v_{\rm c}$,
which tends to decrease at large corrugations.
The obvious consequence is a change of paradigm in tribology, where semi-classical newtonian models need to be replaced by quantum
approaches. Within the CNT, $^4$He does not overcome potential barriers in the classical Newtonian sense: due to its wave-like
character it undergoes effective diffraction by the periodic interface potential.

Scattering from impurities is commonly associated with finite friction forces even at $T$=0 K. However, in realistic (5,5)CNTs the
$^4$He-impurity interactions are associated with mean free paths that exceed the $\mu$m scale. Additionally, impurity contributions
can be experimentally suppressed by CVD growth and clean environments, becoming essentially negligible.
On the other hand, impurity scattering rates are expected to rapidly increase when the
CNT diameter gets larger, so that near-1D confinement turns out to be a key factor in ballistic transport.
In the same breath, thermally excited phonons/plasmons could potentially introduce additional scattering channels at $T>$0 K.
Surprisingly, these contributions have even smaller impact due to weak coupling, especially close to 1BZ edges.
Overall, ballistic transport remains possible even under realistic conditions.
CNT translations are also implicit in our model, and can occur as a consequence of He scattering against
the interface potential or defects/impurities due to conservation of the total momentum (see also Supplemental Material, Ref.~\cite{SM}).

In its present form, our model is not meant to describe many-body effects or complex internal
degrees of freedom, that are expected to be relevant in near-1D water flow.
The limit of individual $^4$He atoms considered in this work experimentally corresponds to a viable low-density regime.
When $^4$He at low density is pumped through a CNT by a minimal pressure gradient, weak $^4$He-$^4$He vdW interactions are expected to
enable large inter-atomic separations and minimization of non-additive many-body effects.

We finally note that while the model Eq.~\eqref{model1d} was meant to describe $^4$He flow through (5,5) CNTs, the equations derived in this work are easily extendable to different flowing atoms/molecules and to alternative nanotubes. The generality of Eqs.~\eqref{system} further suggests possible applicability of our mechanism even beyond 1D. Relations with superlubricity phenomena in 2D nanomaterials also deserve accurate investigation.

\section{Conclusions}
We found that ballistic transport --a quantum phenomenon conventionally associated with unimpeded electron transport in near-1D systems--  is extendable
to atomic $^4$He flow through narrow CNTs. This result changes the general perspective on ballistic phenomena, and could help interpret the giant permeability enhancements observed when water flows through CNTs approaching nm-scale diameters.  The key prerequisites for ballistic atomic flow are weak $^4$He-CNT interactions, near-1D confinement (or small CNT diameter), and low concentration of lattice defects in the CNT - a condition which is easily achieved at the experimental level.
We also note that
ballistic atomic/molecular transport extends near-unimpeded mass flow - a phenomenon that was only associated with superfluidity and
ultra-cold temperatures - to new regimes, characterized by much higher temperatures (300 K), and velocities
(up to $\sim$ 200 m/s).
Ballistic atomic transport can thus profoundly transform nanofluidics and green technologies, providing unparalleled energy-efficiency in
fluid transport and filtration. Moreover, the deep quantum nature of flowing $^4$He opens the way to novel
interference phenomena and quantum coherence, in parallel with ballistic electrons.

\section{Methods} DFT provides quantum mechanical account for the electronic structure of the whole system.
The semi-local Perdew-Burke-Ernzerhof~\cite{PBE} (PBE) functional is adopted to account for electronic
exchange and correlation terms,
in combination with Grimme's~\cite{d2} D2 van der Waals~\cite{rsscs,proof,exact} (vdW) corrections.
The Quantum Espresso~\cite{qe} suite was exploited for all DFT calculations, introducing a periodic description of the system
based on super-cells and ultrasoft pseudopotentials. Electronic wavefunctions are expanded on plane-waves, with an
energy cutoff which
was fixed to 35 Ry. Simulation super-cells contain 8 longitudinal replicas of the unit cell for armchair
nanotubes, and 4 replicas for the larger zigzag nanotubes, while a lateral vacuum space of 15 \AA\,
separates the periodic replicas, orthogonally to the NT axis. Details on analytical calculations are reported in the supplementary material.

\section{Acknowledgements}
A.A., L.S. and P.L.S. are grateful to Prof. Francesco Ancilotto for fruitful discussion.
A.A. has partly been funded by the European Union - Next-Generation EU (Project CN00000041, CUP B93D21010860004) Mission 4, Component 2, Investment 1.4 - PNRR - National Center for Gene Therapy and Drugs based on RNA Technology.- CN3 - Spoke 7 (Biocompting) CUP C93C22002780006, and by University of Padova, 
PARD grant "Quasi-frictionless water superflow in narrow carbon nanotubes: unraveling the true quantum mechanical
mechanism". P.L.S. and L.S. are partially supported by the European Quantum
Flagship Project "PASQuanS 2", by the European Union-NextGenerationEU within the National Center for HPC, Big Data and Quantum Computing
(Project No. CN00000013, CN1 Spoke 10: “Quantum Computing”). L.S. is also supported by the BIRD Project "Ultracold atoms in curved geometries" of the
University of Padova, and by “Iniziativa Specifica Quantum” of INFN. L.S. also thanks the Italian Ministry of University and Research for the PRIN grant "Quantum Atomic Mixtures: Droplets, Topological Structures, and Vortices". A.A. and L.S. acknowledge the Italian Ministry of University and Research for the grant "Dipartimenti di Eccellenza - Frontiere Quantistiche".

\end{document}